\documentstyle[editedvolume,epsfig,amssymb]{crckapc} 
\makeatletter

\def\thebibliography#1{\section*{References\@mkboth
  {REFERENCES}{REFERENCES}}\list
  {\relax}{\setlength{\labelsep}{0em}
        \setlength{\itemindent}{-\bibhang}
        \setlength{\itemsep}{0pt}
        \setlength{\parsep}{0pt}
        \setlength{\leftmargin}{\bibhang}}
    \def\newblock{\hskip .11em plus .33em minus .07em}
    \sloppy\clubpenalty4000\widowpenalty4000
    \sfcode`\.=1000\relax}

\newlength{\bibhang}
\let\@internalcite\cite
\def\cite{\let\@citeleft(\let\@citeright)%
    \@ifstar{\citeyear}{\citefull}}
\def\citenp{\let\@citeleft\relax\let\@citeright\relax
    \@ifstar{\citeyear}{\citefull}}
\def\citefull{\def\astroncite##1##2{##1~##2}\@internalcite}
\def\citeyear{\def\astroncite##1##2{##2}\@internalcite}

\def\@citex[#1]#2{\if@filesw\immediate\write\@auxout{\string\citation{#2}}\fi
  \def\@citea{}\@cite{\@for\@citeb:=#2\do
    {\@citea\def\@citea{; }\@ifundefined
       {b@\@citeb}{{\bf ?}\@warning
       {Citation `\@citeb' on page \thepage \space undefined}}%
{\csname b@\@citeb\endcsname}}}{#1}}

\def\@cite#1#2{\@citeleft#1\if@tempswa , #2\fi\@citeright}
\def\@biblabel#1{}

\newcommand{\msun}{$M_\odot$} 
\newcommand{\persec}{\mbox{$\second^{-1}$}}
\newcommand{\percm}{\mbox{$\cm^{-2}$}}
\newcommand{\ppm}{\mbox{$\pm$}}

\newcommand{\cgslum}{\erg\ \persec}

\newcommand{\approxgt}{\mbox{$\gtrsim$}}
\newcommand{\apj}{\mbox{\rm{Ap. J.}}}
\newcommand{\apjl}{\mbox{\rm{Ap. J.}}}
\newcommand{\mnras}{\mbox{\rm{MNRAS}}}
\newcommand{\aap}{\mbox{\rm{A. \& A.}}}
\newcommand{\pasj}{\mbox{\rm {PASJ}}}
\def\etal{{et~al.}}

\def\xsix1608{{4U~1608$-$522}}
\def\ufifteen{4U~1543$-$47}
\def\hseventeen{H~1705$-$25}
\newcommand{\lxlv}{\mbox{$L_x/L_{\rm bol}$}}
\def\x2129{{4U~2129+47}}
\def\cenx4{{Cen~X$-$4}}

\def\nmon{{A0620$-$00}}

\def\qzvul{{GS~2000+25}}
\def\v404cyg{{GS~2023+33}}
\def\nsco{{GRO~J1655$-$40}}

\newcommand{\nh}{\mbox{$N_{\rm H}$}}

\def\aql{{Aql~X$-$1}}

\newcommand{\ud}[2]{\mbox{$^{+ #1}_{- #2}$}}

\newcommand{\ee}[1]{\mbox{$10^{#1}$}}
\newcommand{\tee}[1]{\mbox{$\times 10^{#1}$}}

\newcommand{\keV}{\mbox{$\rm\,keV$}}
\newcommand{\MeV}{\mbox{$\rm\,MeV$}}
\newcommand{\cm}{\mbox{$\rm\,cm$}}
\newcommand{\km}{\mbox{$\rm\,km$}}
\newcommand{\second}{\mbox{$\rm\,s$}}

\newcommand{\erg}{\mbox{$\rm\,erg$}}
\newcommand{\gauss}{\mbox{$\rm\,G$}}

\newcommand{\chandra}{{\em Chandra\/}}
\newcommand{\rosat}{{\em ROSAT\/}}
\newcommand{\asca}{{\em ASCA\/}}
\newcommand{\rxte}{{\em RXTE\/}}
\newcommand{\xmm}{{\em XMM-Newton\/}}
\newcommand{\beppo}{{\em BeppoSAX\/}}

\newcommand{\kteff}{$kT_{\rm eff}$}

\def\apj{Ap.J.}
\def\apjl{Ap.J. (Letters)}
\def\apjs{Ap.J. (Suppl)}
\def\aa{Astronomy \& Astrophysics}
\def\aapr{Astronomy \& Astrophysics Reviews}		  

\def\aj{A.J.}
		  
\def\mnras{M.N.R.A.S.}
\def\nat{Nature}

\def\pasj{Publ.Astron.Soc.Japan}

\def\araa{Ann.Rev.Astr.Ap.}

\begin{opening}
\title{Transient Low-Mass X-Ray Binaries in Quiescence}
\author{Lars Bildsten} 
\institute{Institute for Theoretical Physics and Department of Physics\\
           Kohn Hall, University of California, Santa Barbara\\
           Santa Barbara, CA 93106} 
\author{Robert E. Rutledge}
\institute{Space Radiation Laboratory, MS 220-47\\
           Caltech, Pasadena, CA 91125}
\end{opening}
\runningtitle{Faint Emission from Neutron Stars and Black Holes}
\begin{document}

We summarize the quiescent X-ray observations of transient low-mass
X-ray binaries. These observations show that, in quiescence, binaries
containing black holes are fainter than those containing neutron
stars. This has triggered a number of theoretical ideas about what
causes the quiescent X-ray emission. For black hole binaries, the
options are accretion onto the black hole or coronal emission from the
rapidly rotating stellar companion.  There are more possibilities for
the neutron stars; accretion, thermal emission from the surface or
non-thermal emission from a ``turned-on'' radio pulsar. We review
recent theoretical work on these mechanisms and note where current
observations can distinguish between them. We highlight the
re-analysis of the quiescent neutron star emission by Rutledge and
collaborators that showed thermal emission to be a predominant
contributor in many of these systems. Our knowledge of these binaries
is bound to dramatically improve now that the \chandra\ and \xmm\
satellites are operating successfully,

\section{Introduction}

Many black holes and neutron stars are in binaries where a
steady-state accretion disk (one that supplies matter to the compact
object at the same rate as mass is donated from the Roche-lobe filling
companion) is thermally unstable \cite{jvp96,king96}. This instability
results in a limit cycle -- as in dwarf novae (where the compact
object is a white dwarf) -- with matter accumulating in the outer disk
for months to decades until a thermal instability is reached
\cite{huang89,mineshige89} that triggers rapid accretion onto the
compact object. The substantial brightening in the X-rays (typically
to levels near the Eddington limit, $10^{38}-10^{39} \ {\rm erg \
s^{-1}}$ for $M=1-10 M_\odot$ stars) brings attention to these
otherwise previously unknown binaries. Both neutron stars (NS) and
black holes (BH) exhibit these X-ray outbursts, separated by periods
($\sim$ months to decades) of relative quiescence (for recent reviews
of the outburst properties, see
\citenp{tanakalewin95,tanaka96,chen97}). The neutron stars are
identified by Type I X-ray bursts from unstable thermonuclear burning
on their surfaces. For those that ``appear'' to be black holes (based
on their spectral and/or timing properties and lack of Type I bursts),
detailed optical spectroscopy in quiescence is undertaken. Many of
the measured optical mass functions are in excess of the maximum
possible neutron star mass ($\approx 3 M_\odot$), making these
binaries an excellent hunting ground for black holes (see
\citenp{mcclintock98} for a summary). 

Our purpose is to discuss the X-ray emission from
these binaries when they are in their faint ``quiescent'' state between
outbursts.  X-ray observations with sensitive pointed instruments
(ROSAT and ASCA) of these transients in quiescence have detected all
of those harboring neutron stars and some that contain black holes.
It is clear that, on average, the binaries containing black holes are
less luminous than those with neutron stars
\cite{barret96,narayan97b,asai98}. 
It is still a mystery as to what powers the very faint X-ray emission
($L_x\ll 10^{35} \ {\rm erg \ s^{-1}}$) from these binaries when in
quiescence, and we will review the possibilities here. 

  Accretion is the most often discussed energy source and clearly powers the
dwarf novae (the analogous systems that contain white dwarfs in
systems with orbital periods typically less than three hours) in
quiescence. These were found by {\it Einstein} to be faint X-ray
sources ($10^{30}-10^{32} {\rm erg \ s^{-1}}$) when in their quiescent
state \cite{cordova84,patterson85a}.  The inferred accretion rate onto
the white dwarf is a few percent of the rate being transferred within
the binary, and the X-rays originate from the boundary layer near the
white dwarf \cite{patterson85a}.  This was confirmed via eclipse
observations with ROSAT of three short orbital period DN in quiescence
\cite{mukai97,vanteeseling97,pratt99}.  In all of these systems, the
X-ray emission was eclipsed when the white dwarf was behind the
companion. The physics that sets this low inflow rate towards the
white dwarf is not clear and might well be different than in the
binaries containing neutron stars and black holes, which are the focus
of this review.

\newpage
\section{Black Hole Transients in Quiescence: Advection Dominated
Accretion Flows  or Coronal Emission?}

  At this time, three BHs have been detected in quiescence: \nmon\
\cite{mcclintock95}, \v404cyg\ \cite{verbunt94,wagner94}, and \nsco\
\cite{hameury97}. The puzzle of the emission mechanism began when
ROSAT/PSPC detected X-rays from \nmon\ at a level $L_x\approx 6\times
10^{30} \ {\rm erg \ s^{-1}}$ \cite{mcclintock95}.  McClintock et
al. (1995) made it clear that this X-ray emission could not be due to
a steady-state accretion disk around the black hole, as if so, there
would be a production of optical and UV photons from the outer parts
of the disk that would far exceed that observed. 

  To solve this puzzle and to explain the higher quiescent
luminosities ($10^{32}-10^{33}\ {\rm erg \ s^{-1}}$) of the transient
NSs, Narayan et al. \cite*{narayan97a} invoked an advection-dominated
accretion flow (ADAF) onto the compact object at a rate $\dot M_q$ in
quiescence. For the black holes, the X-rays are produced
via Compton up-scattering of the optical/UV synchrotron emission from
the inner parts of the flow. The model predicts X-ray emission as a
fraction of the optical/UV emission (in excess of that from the
stellar companion) -- an observable ratio which is used to evaluate
the model's success. Current ADAF spectral modeling of the X-ray
detected BH's requires that $\dot M_q$ be $\sim 1/3$ of the total mass
transfer rate in the binary \cite{narayan97b}.  Accretion rates this
high are required because of the relative inefficiency of the flow at
producing X-rays (\citenp{narayan97a,hameury97}). The much higher
efficiency (by 3-4 orders of magnitude) of X-ray production from
accretion onto a neutron star forces the quiescent accretion rate onto
these objects to be much lower than for the black holes. In other
words, if the implied $\dot M_q$ from BHs was landing on a NS, it
would shine in quiescence at about $10^{36} \ {\rm erg \ s^{-1}}$, a 
factor of 1000 brighter than observed. A solution is
to just dial $\dot M_q$ to be uniformly 
lower in the NS systems than in the BH
systems; though this is not easy to do \cite{menou99}.

 Another possible mechanism for the faint emission from black hole
binaries is coronal X-ray emission from the tidally locked companion
star \cite{verbunt96,bildsten00}.  The analogous systems are tidally
locked stars in tight binaries, such as the RS~CVn systems. These have
X-ray luminosities from coronal activity that reach the level observed
from the black hole transients. For a convective star that is rotating
rapidly, most X-ray observations point to $L_x/L_{\rm bol} \approx 
10^{-3}$ as a ``saturation limit'' in coronal X-ray emission
\cite{vilhu87,singh99}.

\begin{figure}[h]
\centerline{\epsfig{figure=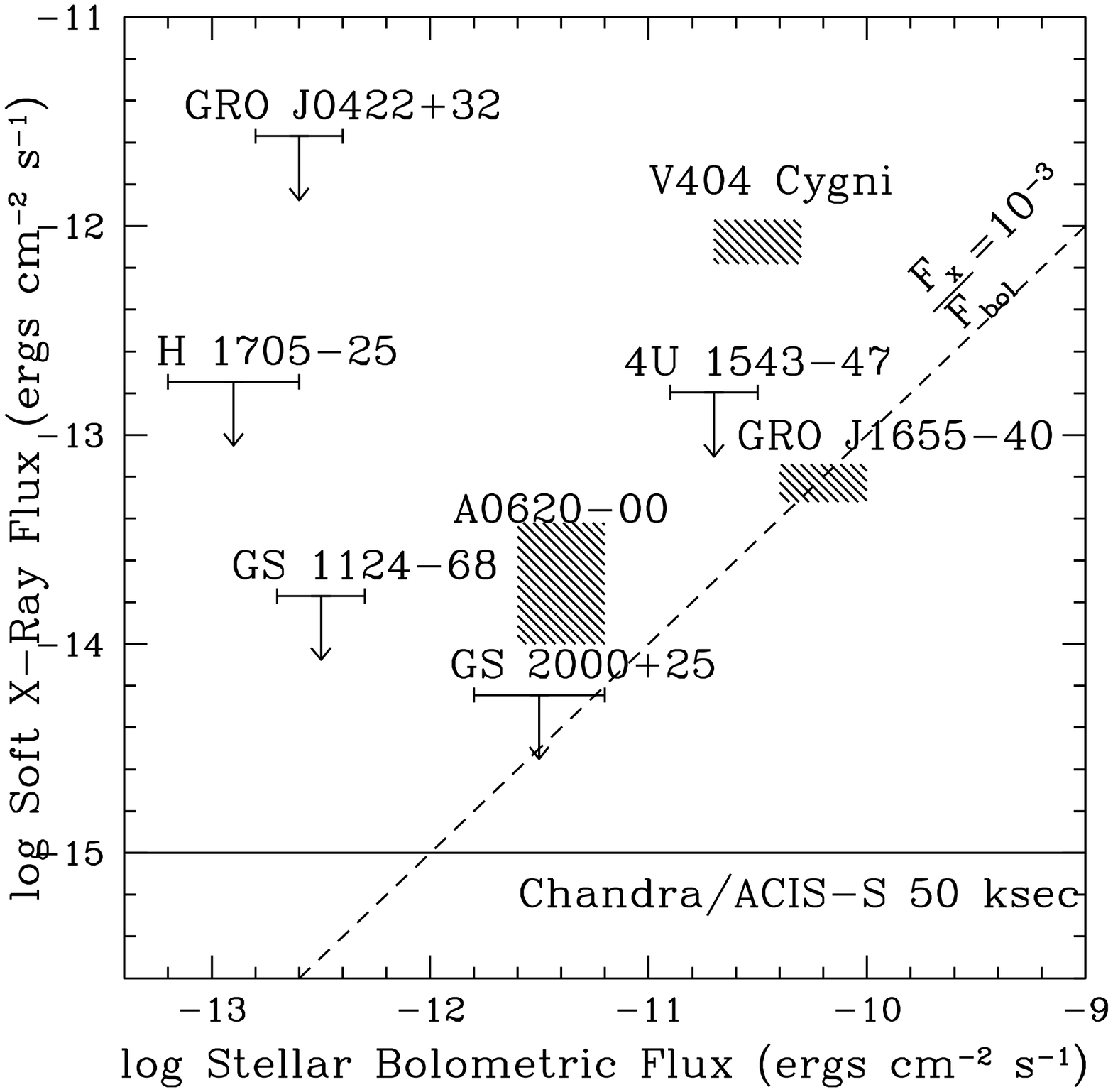,height=7cm,,width=7cm,angle=0}}
\caption[]{ X-ray flux vs. stellar bolometric flux for X-ray detected
black hole binaries (shaded regions) and 2$\sigma$ upper-limits. The
dashed line shows a typical coronal value $F_x/F_{\rm bol}$=$10^{-3}$.
The bottom line is the \chandra/ACIS-S 50 ksec detection limits for a
$N_{\rm H}$=0.2\tee{22} \percm, Raymond-Smith model at 
$kT$=1.0 keV. Detection of X-rays from GRO~J0422+32,
GS~1124-68, and \hseventeen\ with \chandra\ would be well above that 
expected from coronal emission. \qzvul\ is close to the
\lxlv=$10^{-3}$ limit and may be detected.  While \ufifteen\ is
apparently well within X-ray detectable range, the early-type
companion (A2V) is non-convecting, which makes it a ``clean'' system
to study X-ray emission which is not coronal. From Bildsten \&
Rutledge (2000).} \label{fig:corona}
\end{figure}

In Fig.~\ref{fig:corona}, we display the X-ray detections and
upper-limits for several observed BH systems, with their
optically-derived bolometric flux, along with this saturation limit in
\lxlv, and the prospects of detecting coronal X-rays from undetected
systems with \chandra \ (X-ray flux limits for \xmm\ are about a factor
three of lower).  Of the previously undetected sources, only \qzvul\
stands out as a possible new detection of stellar coronal emission
(the companion of \ufifteen\ is not convective, and thus no coronal
emission is expected).

  Distinguishing between these two competing mechanisms -- ADAFs and
coronal emission -- can be done with high S/N X-ray spectroscopy
across the 0.1-4.0 keV energy range, where the X-ray emission is
detected. Stellar coronal X-ray spectra are measured in low-mass stars
and are reasonably well described as a Raymond-Smith plasma
\cite{raymondsmith}. These X-ray spectra are distinct from those
calculated from ADAFs, which consist of a featureless continuum from
the Compton-scattered optical/UV emission and weak line emission that
does not appear detectable with the current generation of detectors
\cite{narayan99}.

\section{X-Ray Spectra of Quiescent Neutron Stars}

Centaurus X-4 was the first NS transient detected in quiescence
\cite{jvp87}.  More recently, quiescent X-ray spectral measurements
have been made of \aql\ \cite{verbunt94} and 4U~2129+47
\cite{garcia99} with the \rosat/PSPC; of EXO~0748$-$676 with Einstein
IPC \cite{garcia99}; and of \cenx4\ and \xsix1608\ with \asca\
\cite{asai96b}. The X-ray spectrum of \aql\ (0.4--2.4\keV) was
consistent with a blackbody (BB), a bremsstrahlung spectrum,
or a pure power-law \cite{verbunt94}.  For \xsix1608, the
spectrum (0.5--10.0\keV) was consistent with a BB ($kT_{\rm BB}\approx
0.2\mbox{--}0.3\keV$), a thermal Raymond-Smith model ($kT =
0.32\ud{0.18}{0.5}\keV$), or a very steep power-law (photon index
$6\ud{1}{2}$). Similar observations of \cenx4\ with \asca\ found its
X-ray spectrum consistent with these same models, but with an
additional power-law component (photon index $\approx2.0$) above
5.0\keV\ (recent observations with \beppo\ of \aql\ in quiescence also
revealed a power-law tail; \citenp{campana98a}).  These high energy
power-law components are not fully understood, and their relationship
to the thermal component is unclear. We discuss this more in \S 4.3. 

In four of these five sources (the exception being EXO~0748$-$676), BB
fits implied an emission area with a radius $\approx 1\km$, much smaller
than a NS. This has little physical meaning however, as the emitted
spectrum from a quiescent NS atmosphere with light elements at the
photosphere is far from a blackbody \cite{romani87}.  For a weakly-magnetic
($B\leq10^{10}$G) pure hydrogen or helium \label{sec:foot}
\footnote{The strong surface gravity will stratify the atmosphere 
within $\sim 10\second$ \cite{alcock80,romani87}. Hence, for
accretion rates $\lesssim 2\times 10^{-13}M_\odot {\rm\,yr^{-1}}$
(corresponding to an accretion luminosity $\lesssim 2\times
10^{33}\cgslum$), metals will settle out of the photosphere faster
than the accretion flow can supply them (Bildsten, Salpeter, \&
Wasserman \cite*{bildsten92}).  As a result, the photosphere should
be nearly pure hydrogen if $\dot M_q$ is small.} 
atmosphere at effective temperatures
\kteff $\lesssim0.5\keV$ the opacity is dominated by free-free
transitions \cite{rajagopal96,zavlin96}.  Because of the opacity's
strong frequency dependence ($\propto\nu^{-3}$), higher energy photons
escape from deeper in the photosphere, where $T>T_{\rm eff}$
\cite{pavlov78,romani87,zampieri95}.  Spectral fits near the peak and
into the Wien tail
(which is the only part of the spectrum sampled with current
instruments) with a BB curve then overestimate $T_{\rm eff}$ and
underestimate the emitting area, by as much as two orders of magnitude
\cite{rajagopal96,zavlin96}.

\begin{figure}[h]
\centerline{\epsfig{figure=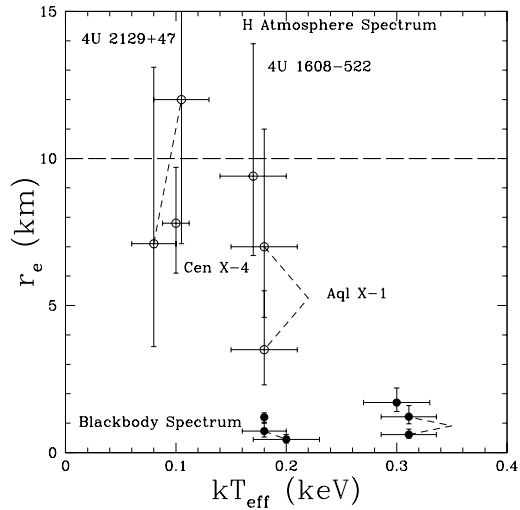,height=7cm,width=7cm,angle=0}}
\caption[]{ Comparison between the spectral parameters $r_e$ and
\kteff, derived from spectral fits of the quiescent X-ray emission
from \aql, \cenx4, \xsix1608 and \x2129.  The {\em open points} are
for the H atmosphere spectrum and the {\em solid points} are from a
black-body spectrum.  The two points connected for \aql\ correspond to
the upper- and lower- distance limits for that source.  The two
connected points for \x2129\ are for two different distance/\nh\
estimates.  The H atmosphere fits produce values of $r_e$, consistent
with a 10~km NS. From Rutledge \etal\ (2000).  }
\label{fig:nsbbhatm}
\end{figure}

Rutledge et al. \cite*{rutledge99,rutledge00} showed that fitting the
spectra of quiescent NS transients with realistic atmospheric 
 models yielded emitting
areas consistent with a 10~km radius NS.  In Fig.~\ref{fig:nsbbhatm},
we compare the measured H atmosphere and blackbody spectral parameters
for the quiescent NSs. The emission area radii are larger from the H
atmosphere spectra by a factor of a few to ten, and are consistent
with the canonical radius of a NS. There is thus observational
evidence that thermal emission from a pure hydrogen photosphere
contributes to -- and perhaps dominates -- the NS luminosity at photon
energies of 0.1--1 keV. This will be tested much better with upcoming
{\it Chandra} and {\it XMM-Newton} observations. 

\section{What Powers the Quiescent Emission from the Neutron Stars?}

 In our earlier discussion of black holes, we pointed out that the
expected coronal X-ray emission from the tidally locked and rapidly
rotating stellar companions is at a level consistent with that
observed from the black hole transients A~0620-00 and GRO
J1655-40 (though see \citenp{lasota00} for an argument against this).
The BH transient V404 Cygni is too bright to be explained
this way, as are all binaries containing NSs.
Several energy sources for the quiescent NS emission have been
discussed and developed \cite{stella94}. These include 
late-time thermal emission from heat released deep in the
NS crust during outbursts \cite{brown98}, accretion
\cite{jvp87,menou99}, and non-thermal emission from a turned-on
radio pulsar \cite{campana98b}.

\subsection{Thermal Emission from Deep Nuclear Energy Release}

Brown, Bildsten and Rutledge (1998) showed that the ``rock-bottom''
emission from these systems is set by thermal emission from the
neutron star. This minimum luminosity comes from nuclear energy
deposited in the inner crust (at a depth of $\approx 300 \ {\rm m}$)
during the
large accretion events. The freshly accreted material compresses the
inner crust and triggers nuclear reactions that deposit about an MeV 
per accreted baryon there \cite{haensel90}. This heats the NS
core on a \ee{4}-\ee{5} yr timescale, until it reaches a steady-state
temperature $\approx 4\times10^7
(\langle\dot{M}\rangle/10^{-11}\,M_\odot{\rm\,yr^{-1}})^{0.4}\rm\,K$
\cite{bildsten97}, where $\langle\dot{M}\rangle$ is the time-averaged
accretion rate in the binary. 
A core this hot makes the NS ``glow'' at a luminosity
\begin{equation}
\label{eq:lq}
L_q \approx 
{1\MeV\langle\dot{M}\rangle\over m_p}
\approx 6\times 10^{32}{{\rm erg}\over {\rm s}}
\left(\langle\dot{M}\rangle\over 10^{-11}M_\odot\rm\,yr^{-1}\right),
\end{equation} 
even after accretion halts \cite{brown98}. The NS is then a thermal
emitter in quiescence, consistent with the inferrences from the
quiescent spectroscopy noted earlier.  This quiescent emission is
inevitable (unless accelerated core cooling mechanisms are active in
these stars that do not seem to occur in young neutron stars) and
provides a ``floor'' for the quiescent luminosity. Additional emission
mechanisms can add to this, increasing the luminosity and modifying
the spectral shape. 

\begin{figure}[h]
\centerline{\epsfig{figure=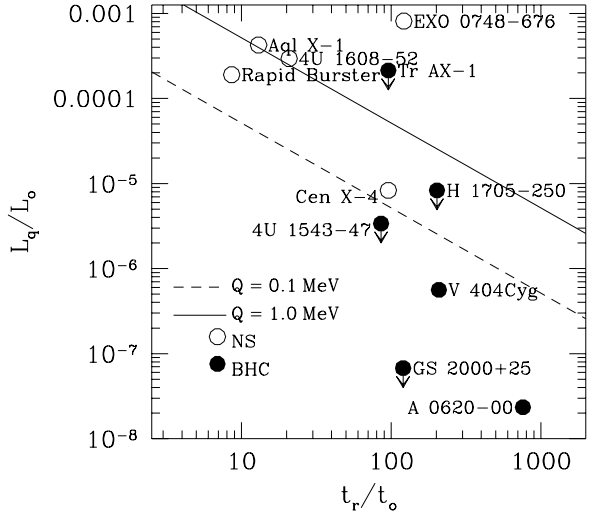,height=7cm,width=7cm,angle=0}}
\caption[]{The ratio of quiescent luminosity $L_q$ to outburst
luminosity $L_o$ as a function of the ratio of recurrence interval
$t_r$ to outburst duration $t_o$.  The lines are for different amounts
of heat, $0.1\MeV$ ({\em dashed line\/}) and $1.0\MeV$ ({\em solid
line\/}), per accreted nucleon deposited at depths where the thermal
time is longer than the outburst recurrence time.  Also plotted are
the observed ratios for several NSs ({\em open circles\/}) and BHs
({\em filled circles}).  For most of the BHs, only an upper limit ({\em
arrow\/}) to $L_q$ is known.  Data is from \protect\cite{chen97}, with
the exception of $L_q$ for the Rapid Burster \protect\cite{asai96b}.
For Aql~X-1 and the Rapid Burster, $L_o$ and $t_o$ are accurately
known (\rxte/All-Sky Monitor public data); for the remaining sources
$L_o$ and $t_o$ are estimated from the peak luminosities and the rise
and decay timescales. From Rutledge \etal\ (2000).  } \label{fig:Lq}
\end{figure}

In exploring this scenario Rutledge et al. (2000) analysed only
observations made during periods of the lowest observed flux, to
minimize contributions from accretion. They calculated the bolometric
luminosity from the H atmosphere fits.  Using these new bolometric
quiescent luminosities for Aql~X-1, Cen~X-4, and 4U~1608$-$522, they
plotted (Fig.~\ref{fig:Lq}) $L_q/L_o$ as a function of $t_r/t_o$
\cite{brown98}.  Here $L_q$ and $L_o$ are the observed quiescent and
average outburst luminosities, and $t_r$ and $t_o$ are the recurrence
interval and outburst duration.  We show this relation for the NSs
({\em open circles\/}) Aql X-1, Cen X-4, 4U~1608$-$522, and
EXO~0748$-$676 and the BHs ({\em filled circles}) H~1705$-$250,
4U~1543$-$47, Tra~X-1, V~404~Cyg (GS~2023+33), GS~2000+25, and
A~0620$-$00.  We denote with an arrow those BHs for which only an
upper limit on $L_q$ is known.

The expected incandescent luminosity is plotted for two 
amounts of heat per accreted baryon deposited in the 
inner crust during an
outburst: 1\MeV ({\em solid line}) and 0.1\MeV ({\em dotted line}).
With the exception of Aql~X-1, Cen~X-4, 4U~1608$-$522, and the Rapid
Burster, 
the data from this plot is taken from \cite{chen97}.  For Aql~X-1 and
the Rapid Burster, $L_o$ and $t_o$ are accurately known (\rxte/All-Sky
Monitor public data); for the remaining sources $L_o$ and $t_o$ are
estimated from the peak luminosities and the rise and decay
timescales.

Four of the five NSs are within the band where the quiescent
luminosity is that expected when the emitted heat is between 0.1-1.0
MeV per accreted baryon.  The fifth NS (EXO~0748$-$676), has a higher
quiescent luminosity (by a factor of 10), which we interpret as being
due to continued accretion, an interpretation which is reinforced by
the observation of spectral variability during the quiescent
observations with {\it ASCA} \cite{corbet94,thomas97}, on timescales
of $\sim$1000 sec and longer. (\citenp{garcia99} measured
$L_x$=1\tee{34} \cgslum\ from Einstein/IPC observations of this
source.)  The BHs on this figure are more spread out across the
parameter-space, qualitatively indicating a statistical difference --
although not one which is particular for each object -- between the
two classes of objects.  This suggests the NS quiescent luminosity is
more strongly related to the accreted energy than the BH quiescent
luminosities.

\subsection{Accretion onto the Neutron Star During Quiescence} 

Accretion was initially suggested \cite{jvp87} as the energy source of
quiescent emission from transiently accreting NSs, partially because
few other emission mechanisms were known at the time. Thermal emission
was not considered, as it was presumed that these neutron stars with
low-mass companions were clearly older than the NS core cooling
timescale ($10^6$ yr) and would have cold cores. The work of Brown et
al. (1998) changed that. 

There are presently no observational results which exclude the
possibility that part of the quiescent luminosity of these NSs is due
to accretion.  Indeed, some observational evidence suggests that
accretion occurs onto the NS surface during quiescence; long-term
(months-years) variability in the observed flux has been reported (a
factor of 4.2\ppm0.5 in 8 days from \cenx4; \citenp{campana97}) and in
4U~2129+47, by a factor of 3.4\ppm0.6 between Nov-Dec 1992 and March
1994; \citenp{garcia99,rutledge00}). If so, then the required rate is
about $\dot M_q=$\ee{-14}--\ee{-15} \msun\ yr$^{-1}$, a factor of
$10^{-4}$ below the time-averaged rate. While this
intensity variability can be explained by a variable absorption column
depth, active accretion during quiescence is also a possibility.

Recent observations of \aql\ at the end of an outburst showed an
abrupt fading into quiescence \cite{campana98a} associated with a
sudden spectral hardening \cite{zhang98a}. This was followed by a
period of $\sim15$ days, over which the source was observed (three
times) with a constant flux level \cite{campana98a}.  This behavior
was interpreted as the onset of the ``propellor effect''
\cite{ill75,stella86} in this object, which would inhibit -- perhaps
completely -- accretion onto the NS.  The energy source
for the long-term nearly constant flux is most likely thermal emission
\cite{brown98}. 

  A thermal spectrum alone cannot distinguish between accretion and a
hot NS core as the energy source. This is because the accretion energy
is likely deposited deep beneath the photosphere and is re-radiated as
thermal emission \cite{zampieri95}.  This emission is nearly identical
to that expected from the hot NS core. The only possible difference
would be if the accretion rate is high enough (about $>10^{-13}
M_\odot \ {\rm yr^{-1}}$, see footnote in \S~\ref{sec:foot}), to
constantly replenish metals in the photosphere and if spallation of
these elements is not too strong \cite{bildsten92}. These metals,
particularly Oxygen, will imprint photoabsorption edges in the
emergent spectrum \cite{rajagopal96}.  The presence of such metallic
absorption in the NS quiescent emission spectra -- aside from being
astrophysically important -- would clearly indicate active accretion
onto the NS.
\label{sec:lines}

\subsection{Magnetospheres and Spins} 

 Evolutionary scenarios that connect the accreting neutron stars in
Low-Mass X-ray Binaries to millisecond radio pulsars predict that
these neutron stars should be rapidly rotating (at a few milliseconds)
and magnetized at $10^8-10^9 \ {\rm G}$.  There is only one
transiently accreting neutron star that unambiguously looks like this,
SAX J1808.4-3658, at $\nu_s=401 \ {\rm Hz}$ \cite{rudi98,deepto98}, 
and $B=10^8-10^9 \ {\rm G}$ \cite{psaltis99}. We know little about the
magnetic fields and spins of the neutron stars in the other
transients.  The only one for which we know the spin is Aql X-1, where
nearly coherent oscillations during Type I bursts imply a rotation
rate of 550 Hz \cite{zhang98b}.

  The neutron star's spin and magnetic field are important for two
reasons. The first is the distinct possibility of shutting off
accretion onto the neutron star from the ``propellor'' effect. This
can happen when the magnetospheric radius exceeds the co-rotation
radius (where the Kepler period equals the spin period), which for a
neutron star spinning at $\nu_s$ with magnetic moment $\mu$, will
happen when the accretion rate is below $\dot M_p\approx 7\times
10^{-11} M_\odot \ {\rm yr^{-1}}(\mu/10^{26} {\rm G \ cm^{3}})^2
(\nu_s/300 \ {\rm Hz})^{7/3}$, suggesting a minimum accretion
luminosity of $\approx 10^{36} \ {\rm erg \ s^{-1}}$ for the fiducial
parameters in the accretion rate equation. Even once in this regime,
Campana \etal\ \cite*{campana98b} discussed the possibility of
emission from the gravitational energy release of matter striking the
magnetosphere itself, which would yield a lesser amount of energy per
gram and thus a lower total luminosity.  Finally, if a magnetic field
plays an important role in the geometry of quiescent accretion, one
might expect some asymmetries that would produce X-ray
pulsations. This was not observed in the recent fading of Aql~X-1
\cite{campana98a,zhang98a}, where stringent limits on the pulsed
fraction of the emission were placed ($\leq$1.2\% rms variability,
95\% confidence; \citenp{chandler00}) at a time believed to be just at
the onset of the propellor for a $10^8 {\rm G}$ field. What this implies
about the magnetic field strength of Aql X-1 is still unknown.

The second place where the magnetic field and spin matter is when the
magnetosphere becomes larger than the light cylinder. One might imagine
the neutron star turning into a millisecond radio pulsar at this stage
(see \citenp{stella94} for an overview). However, a millisecond radio
pulsar at the position of a transient X-ray binary has never been
observed, even for SAX J1808.4-3658. Perhaps it is difficult for the
accretion rates in these systems to become low enough to allow pulsar
activity.

 Campana \etal\ \cite*{campana98a} conjectured that the hard X-rays
sometimes seen in quiescence might be non-thermal emission from an
active pulsar.  If so, then the energy source is a fraction of the
spin-down luminosity. As Stella et al. (1994) noted, if the fraction
of spin-down energy going into X-rays in transiently accreting
binaries is similar to that observed from millisecond radio pulsars
($L_x\approx 10^{-3} \dot E$; Becker \& Tr\"umper 1999), and the
magnetic field strengths are sufficient to ``turn on'' a millisecond
pulsar, then the predicted X-ray luminosities are close to those
observed ($\sim$\ee{32}-\ee{33} \cgslum). Indeed, a few of the X-ray
detected millisecond pulsars have X-ray luminosities in this range
(Becker \& Tr\"umper 1999).  Brown et al. (1998) noted the same
possibility for the neutron star in SAX J1808.4-3658. The X-ray
spectral energy distribution of such a non-thermal component is hard
and power-law like (Becker \& Tr\"umper 1999) 
similar to the hard power-law tails observed in
Cen X-4 and Aql X-1 \cite{asai96b,campana98a}.

\section{Conclusions and The Future} 

It is only recently that the focus of studying quiescent NSs has
changed from parameterizing the phenomenology, to measuring the
physics behind the emission.  Higher quality X-ray data from the
X-ray spectroscopy missions \chandra and  \xmm  \ will
provide much better data than any of the previous missions.  These
will also provide the means to account for possible contributions due
to a hard-power law component in the black holes, as well as the neutron
stars.

Bildsten and Rutledge \cite*{bildsten00} have recently argued that two
(\nmon \ and \nsco) of the three X-ray detected black hole binaries
exhibit X-ray fluxes entirely consistent with coronal emission from
the companion star. The current upper limits on the remaining BHs are
also consistent with production via chromospheric activity in the
secondary.  All four NSs (\aql, \cenx4, \xsix1608, \x2129) have
quiescent X-ray luminosities which are at least ten times greater than
expected from chromospheric emission alone.

This suggests that a viable hypothesis for the majority of the
transient NSs and BHs is that little accretion occurs in
quiescence. Though mass is continuously transferred from the companion
to the outer accretion disk (as is clear from the H$\alpha$ line
emission), accretion onto the compact object appears to be small. In
the absence of accretion, the quiescent X-rays from a NS would then be
dominated by thermal flux from a hot NS core \cite{brown98}, while for
BHs, the quiescent X-rays come from the chromospheric activity of the
secondary.  The advantage of this hypothesis is that it explains the
X-ray luminosities of BHs and (separately) NSs, without having to
invoke dramatically different quiescent accretion rates 
that depend on the type of compact object \cite{menou99}.  

  At odds with this simple scenario is the detection of X-rays from
V404 Cygni at a level which is a factor of ten brighter than can be
explained as coronal emission.  In addition, the observed variability
in the NS quiescent luminosity can not be explained easily without
some accretion.  Both of these observations point to the possibility
that the transferred matter sometimes can make it down to the central
compact object.  The fraction of the time this occurs and the reason
why still needs to be better understood.

One of the most important areas of research in the coming years will
be the search for and exploitation of photospheric absorption edges in
NS quiescent spectra (\S~\ref{sec:lines}). These edges are a kind of
``holy grail'' of NS spectroscopy as the known energy permits us to
measure the gravitational redshift.  We can also use realistic
atmospheric spectra to derive the emission area radius divided by the
distance to the NS.  This radius, combined with the photospheric
redshift, will provide an independent measure of the NS mass and
radius, and thus its equation of state. The major uncertain parameter
in these systems -- the source distance -- can be measured with the
Space Interferometric Mission, set for launch in 2006, which will
measure parallactic distances to objects as faint as 20th magnitude to
4$\mu$arcsec; which can find the majority of systems in this review.

In addition to better understanding of known sources, we hope that the
new satellites will also probe quiescent emission from other populations
of transient accretors.  A likely place for progress with \chandra\
are the low-luminosity X-ray sources observed in globular clusters
\cite{hertz83} which are either cataclysmic variables
\cite{cool95,grindlay95} or transient neutron stars in quiescence
\cite{verbunt84}.  X-ray spectroscopy can identify these objects as
NSs radiating thermal emission from the atmosphere, or imply a
different origin for the emission. As discussed above and elsewhere
\cite{brown98}, the quiescent luminosities of these sources are set by
the time averaged accretion rate. Thus, the low luminosity ($10^{31}$
\cgslum) X-ray sources in globular clusters, if they were transient
neutron stars in quiescence, would have $\langle \dot{M} \rangle
\approx 2\tee{-13}$ \msun $\ {\rm yr}^{-1}$.  The advantage to the
cluster work will be the prior knowledge of the distance and reddening
to the sources. {\it Simultaneous thermal spectroscopy of multiple
sources in the same globular cluster at a known distance might well
provide the first unambiguous and simultaneous 
measurements of many neutron star radii!}

We also expect progress in quiescent observations of transiently
accreting X-ray pulsars.  These systems are typically in high mass
(\approxgt 10\msun) X-ray binaries, where the companion can contribute
a significant fraction of the expected persistent X-ray luminosity,
forcing us to depend on a pulse for secure detection of thermal
emission in quiescence.  The high magnetic fields
($10^{12}\mbox{--}10^{13}$ \gauss) perturb the opacity of the NS
atmosphere and produce a pulse even if the underlying flux is uniform.
Pulsations at the same luminosity level (\ee{32} \cgslum;
cf.~Eq.~\ref{eq:lq}) as observed from the low-magnetic field systems
was recently seen from
A~0535+26 at a time when the circumstellar disc
was absent \cite{negueruela00}. 
 After excluding a magnetospheric origin, this pulsed
emission was interpreted as due either to matter leaking onto the
polar caps or to thermal emission from the NS core \cite{negueruela00}
-- heated from nuclear emission deposited in the inner crust during
the accretion outbursts, as described by Brown \etal\ (1998).

In binaries containing black holes, the major observational challenge
is to distinguish between the ADAF and stellar coronal emission
models.  The best test is likely to be high S/N X-ray spectroscopy in
the 0.1-4 keV range, where spectral lines contribute significantly in
Raymond-Smith plasma (coronal) models, but not in ADAFs
\cite{narayan99}. Once this is done, more focused studies of the
accretion flows around black holes can be carried out.

  We are clearly at the forefront of discovery regarding the physics
of the quiescent emission of neutron stars and black holes. We have
moved beyond initial detection, and at present a variety of mechanisms
have been proposed to explain emission from both black holes and
neutron stars.  In the present era of \chandra\ and \xmm, we will
study these emission mechanisms in detail for the brightest of
sources. The opportunity to detect neutron star photospheric
absorption edges, and the ability to measure the neutron star radius
from the broad-band spectroscopy may well constrain the neutron star
equation of state.

\acknowledgements

 We thank Ed Brown, George Pavlov and Slava Zavlin for the
collaboration on much of this work.  We are grateful to Ed Brown for
preparing Fig.~\ref{fig:Lq}.  This work was supported in part by the
National Science Foundation through Grant NSF94-0174 and NASA via
grant NAG5-3239.  L.B. is a Cottrell Scholar of the Research
Corporation.

\end{document}